# Search for direct reaction products in $^{197}$Au+ $^{20}$Ne from 108 to 171 MeV projectile energy

*Sumana Mukherjee[1], Susanta Lahiri [1,2,3]\*, Sandipan Dasgupta[4], Jagannath Datta[4,5], Sujoy Chatterjee[6], Sharmishtha Bhattacharyya[4,5], Gopal Mukherjee[4,5], Chiranjib Barman[1]*

[1]Department of Physics, Sidho-Kanho-Birsha University, Purulia-723104, India
[2]Department of Chemistry, Diamond Harbour Women's University, South 24 Pargana-743368, India
[3]Department of Chemistry, Rani Rashmoni Green University, Hooghly-712409, India
[4]Physics Group, Variable Energy Cyclotron Centre, Kolkata-70064, India
[5]Homi Bhabha National Institute, Anushaktinagar, Mumbai-400094, India
[6]TLD unit, Variable Energy Cyclotron Centre, Kolkata,70064, India

**Abstract**

The heavy ion induced nuclear reactions are of interest to nuclear physics community since long time. In this paper, we have reported production cross sections of $^{196}$Au, $^{198}$Au, $^{199}$Tl, $^{200}$Tl, $^{201}$Tl, which might be direct reaction (DIR) products, obtained through the interaction of 108-171 MeV $^{20}$Ne on natural gold target. The radioisotopes were identified and were quantitatively measured by off-line gamma spectrometry. We have also used two different simulation codes PACE4 and FLUKA (version 4-3.3). PACE4 did not predicted production of $^{196}$Au and $^{198}$Au, might be because PACE4 does not considers direct reactions, but predicted $^{199}$Tl, $^{200}$Tl, $^{201}$Tl. FLUKA grossly underpredicted production of all these radionuclides.

## 1 Introduction

The heavy ion (HI) induced interactions are either peripheral interactions or central collisions, depending on the degree of overlap between the projectile and target nuclei. Peripheral interactions cause the projectile and target to break into fragments. The resulting residual nuclei carry only modest excitation energies and retain mass numbers, atomic numbers and velocities those are close to the original projectile and target. Target fragmentation ultimately produces spallation products, though fission also becomes significant when dealing with heavy elements. In recent years, nuclear reactions driven by intermediate-energy HIs have drawn significant attention, mainly because the interaction dynamics undergo a notable shift at these energies. On the other hand, central collisions in HI induced nuclear reactions occur at very small impact parameters, where the projectile and target nuclei overlap almost completely, leading to maximal energy deposition and compression of nuclear matter [1, 2].

Reaction of Gold with heavy ions has an extensive prospect of research like, understanding reaction mechanisms, production of neutron deficient isotopes, production of multi-tracers etc. For example, in 1980 Kaufman et al., reported the production of ~28 radioisotopes from the interaction of Au with $^{12}$C and $^{20}$Ne of 4.8- 25-GeV, 7.6-GeV energy, respectively [3]. Lahiri et al., 2002 reported the production of ~54 isotopes with Au and $^{12}$C of 80MeV/A beam energy [4]. Both Kaufman et al. and Lahiri et al. reported qualitatively the production of $^{196}$Au and $^{198}$Au in their experiments, but not provided any quantitative information. The direct reaction products, e.g., $^{196}$Au, $^{198}$Au, $^{199}$Tl, $^{200}$Tl, $^{201}$Tl produced from $^{20}$Ne (252 MeV) or $^{40}$Ar (336 MeV) interactions with gold was quantitatively measured by Irwin Binder [8]. There are other reports on high energy heavy ion interactions with gold targets, but these reports were silent on direct reaction products (Vergani et al., 1993, Liu et al., 1997, Thomas et al. in 1962) [5-7]

In this work, natural self-supported gold foils were irradiated with $^{20}$Ne beam of 108-171 MeV energy with an aim to identify the direct reaction products. In this energy range literature does not provide any data on the direct reaction products produced from $^{20}$Ne and gold interaction.

## 2 Simulation by Monte Carlo codes

Several nuclear simulation codes are available for assessing the viability of nuclear reactions, including PACE4, FLUKA, TALYS, PHITS, etc. Each of these codes was developed with specific aim for particular applications and was tailor made accordingly. Over the past few decades, PACE4 and FLUKA have been widely employed for predicting HI interactions with target materials. All these three codes are based on Monte Carlo simulation technique that incorporate a variety of physical models to evaluate reaction products, yields, cross sections and radiation dose distributions.

## 2.1 PACE4

The PACE4 code (Projection Angular-momentum Coupled Evaporation) has a wide range of applications, including the calculation of fission barriers, reaction cross sections and recoil velocities, and is particularly well suited for HI induced nuclear reactions [10]. In general, nuclear reactions are classified into three categories: (i) direct interactions (DIR), (ii) pre-equilibrium (PEQ) reactions and (iii) equilibrium (EQ) reactions. PACE4 is based on the EQ reaction mechanisms [11]. In this framework, the de-excitation of the compound nucleus is treated using successive Monte Carlo simulations. The cross sections of evaporation residues are calculated using the Bass formula [12]. Optical model parameters for proton, neutron, and α-particle emission are adopted from the report by Perey and Perey [13].

In the present calculations, the diffuseness of the partial-wave distribution (INPUT) was set to 1. The level density option (FYRST) was set to 0. The parameter BARFAC, which incorporates the modified rotating liquid-drop fission barrier proposed by A. J. Sierk [16], was also set to 0. The ratio of the Fermi-gas level density parameter at the saddle point to that at the ground state was taken as unity, while the parameter FACLA was fixed at 10. The code internally determines the level densities and nuclear masses during the de-excitation process. Fission is treated as a decay mode with an adjustable fission barrier. To maintain consistency in the comparative analysis, all input parameters were kept similar to the experimental conditions.

## 2.2 FLUKA4-3.3

FLUKA (v4-3.3) is a versatile Monte Carlo simulation codes and is widely used for a broad range of applications, including target design, dosimetry calculations, particle–matter interaction studies involving neutrinos, heavy ions and cosmic rays as well as detector design [13, 14]. The code is designed to simulate nearly all possible particle–matter interactions and incorporates distinct physical models tailored to different energy regimes. For hadron-induced inelastic collisions, two models, the Generalised Intra-Nuclear Cascade (GINC) model and the Gribov–Glauber multiple-collision mechanism are employed. In both models, equilibrium processes such as fission, γ-ray de-excitation, Fermi break-up and particle evaporation are included. For nucleus–nucleus interactions over a wide energy range, FLUKA adopts external event generators.

The predictive capability of FLUKA has been extensively validated through benchmark experiments. For example, the study by Choudhury et al. [17] demonstrated excellent agreement between experimental data and FLUKA2011.2x.6 predictions for the interaction of a 1.4 GeV proton beam with lead–bismuth eutectic (LBE) target. Although several studies have been reported on the application of FLUKA for projectile energies exceeding 100 MeV/A, to the best of our knowledge, no investigations have been reported at energies below 10 MeV/A. In the present simulations, an unsuppressed approximation was employed with the number of cascades set to $2\times10^{10}$. Experimental parameters, including beam current, target geometry and irradiation time were reproduced in the simulations to ensure consistency between theoretical and experimental conditions.

## 3 Experimental Set up

Five natural gold foils (99.9% purity; Alfa Aesar) were taken for irradiation. The foil thicknesses ranged from 3 to 4.8 mg cm$^{-2}$ (Table 1). The targets were irradiated with a $^{20}$Ne$^{7+}$ ion beam delivered through a collimator of 10 mm diameter. Throughout the irradiation, the targets were continuously cooled by flow of low-conductivity water. The gold foils were mounted on aluminium target holders and irradiated sequentially with $^{20}$Ne$^{7+}$ beams of incident energies 114, 130, 145, 160, and 174 MeV, accelerated by the Room Temperature Cyclotron (K-130) at the Variable Energy Cyclotron Centre (VECC), Kolkata, India. The experimental irradiation parameters have been summarized in Table 1. The typical beam spot diameter was approximately 0.5 cm. The exit energies of the ions were calculated using the SRIM code [18].

After identifying the produced radioisotopes, the following formulas were used to determine the cross section for the particular radioisotope [17]:

$$\sigma_n(E) = \frac{Y_n}{I_{Pro} N_{tg} (1 - e^{-\lambda_n T})} \quad (1)$$

$$Y_n = Y_0 \times e^{-\lambda_n t} \quad (2)$$

where, $\sigma_n$ is the cross section of the n$^{th}$ residual radionuclide at energy E, $Y_n$ is the activity of the isotope (Bq) as recorded in the gamma spectra, $Y_0$ is the activity at EOB (Bq), $I_{Pro}$ is the beam intensity of projectile (particle s$^{-1}$), $N_{tg}$ is the number of target atoms (atoms cm$^{-2}$), T is the irradiation time (s), $\lambda_n$ is the decay constant (s$^{-1}$) of the particular radioisotope and t is the time from end of bombardment to the data taken (s).

Table 1: Irradiation details of the experiment

| Target | Thickness (mg cm$^{-2}$) | Incident energy (MeV) | Exit energy (MeV) | Energy at midpoint of target (MeV) | Irradiation time (h) | Integrated charge (µC) |
|---|---|---|---|---|---|---|
| $^{nat}$Au | 3.0 | 114 | 103.0 | 108.5 | 5.2 | 663 |
| | 4.8 | 130 | 123.5 | 126.8 | 5.2 | 630 |
| | 3.0 | 145 | 138.8 | 141.9 | 4.0 | 257 |
| | 3.0 | 160 | 154.1 | 157.1 | 4.3 | 244 |
| | 3.0 | 174 | 168.5 | 171.3 | 4.6 | 544 |

Offline time-resolved γ-ray spectra were recorded from 0.5 h after the end of bombardment (EOB) up to 90 d post-EOB. Measurements were performed using a CANBERRA p-type high-purity germanium (HPGe) detector with 40% relative efficiency and an energy resolution of 1.86 keV at 1.33 MeV, operated with appropriate lead shielding. The detection system was coupled to a digital spectrum analyser (DSA-1000, CANBERRA), and spectral analysis was carried out using Genie 2K software. Energy and efficiency calibrations of the detector were performed using a standard $^{152}$Eu source (half-life 13.3 a) with an activity of 4253 Bq on the first day of measurement. The sample-to-detector distance was maintained at 5 cm, a geometry chosen to restrict the counting dead time to below 3%. The half-life of each photo peak was calculated from the decay data.

## 4 Error Calculation

The uncertainties considered in the activity measurement are as follows: (i) calibration of the detector, ≈2%; (ii) error occurred in the measurement from the counting statistics. The inaccuracy resulting from counting statistics varied depending on the quantity of the radioisotope produced, (iii) errors from all other sources such as target length measurement, efficiency calibration and beam intensity on each target have been taken into consideration. Around 10% of the total uncertainty for all the sources in addition to the counting error has been considered in the present study [19].

## 5 Result and discussion

A representative gamma spectrum taken after 47.6 h of EOB has been presented in Figure 1. $^{196}$Au has two intense photo peaks at 333.0 and 355.7 keV. However, $^{76}$Kr ($T_{1/2}$=14.8 h) and $^{97}$Zr ($T_{1/2}$=16.7) have photo-peaks within ± 0.5 keV of the above-mentioned photo peaks of $^{196}$Au. Therefore, $^{196}$Au was confirmed from its decay data. On the other hand, $^{198}$Au has only intense gamma peak at 411.8 keV which is common with $^{198}$Tl ($T_{1/2}$=5.3 h). Subsequent analysis of the decay data confirms the production of $^{198}$Au.

The cross sections have been estimated using formula 1 and presented in figure 2(a) and 2(b). Probably, $^{196}$Au and $^{198}$Au have been produced by direct reaction mechanism, i.e., HI knocked out a spin ½ particle from target and by stripping mechanism the target took a spin ½ particle from projectile respectively.

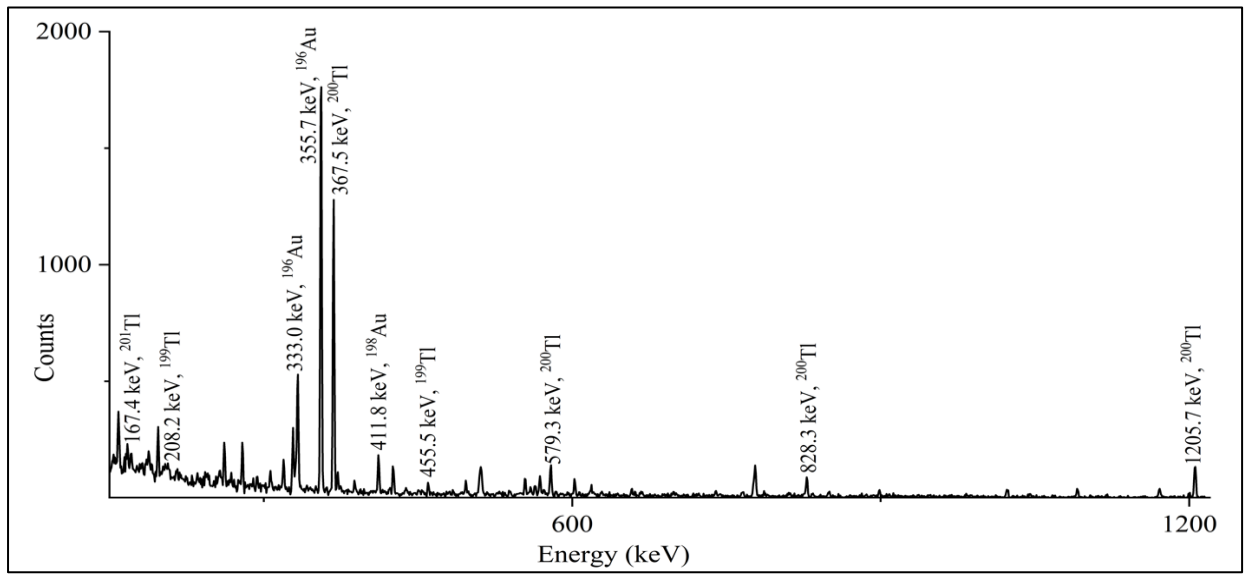

Figure 1: Partial gamma spectrum taken after 47.6 h of EOB, target was bombarded by 169 MeV $^{20}$Ne beam.

Simulation by PACE4 remains silent about the production of $^{196}$Au and $^{198}$Au. PACE4 considers only equilibrium (EQ) reactions and therefore it has not predicted $^{196}$Au and $^{198}$Au which might have produced by direct reactions. Though FLUKA predicted the production of both the gold radioisotopes but grossly underpredicted the production of $^{196}$Au. The experimental excitation functions as well as prediction by FLUKA for $^{196}$Au and $^{198}$Au have been provided in Figure 2(a) and 2(b). The maximum production cross section of $^{196}$Au obtained experimentally was 147.6 mb at 157.1 MeV projectile energy. Although FLUKA underpredicted the cross sections by magnitude, the observed trend was similar to that of experiment. Similarly, excitation functions of $^{198}$Au, both experimental and simulations by FLUKA have been shown in figure 2(b). Interestingly, the simulated cross sections at higher projectile energy are closed to that of experimental one.

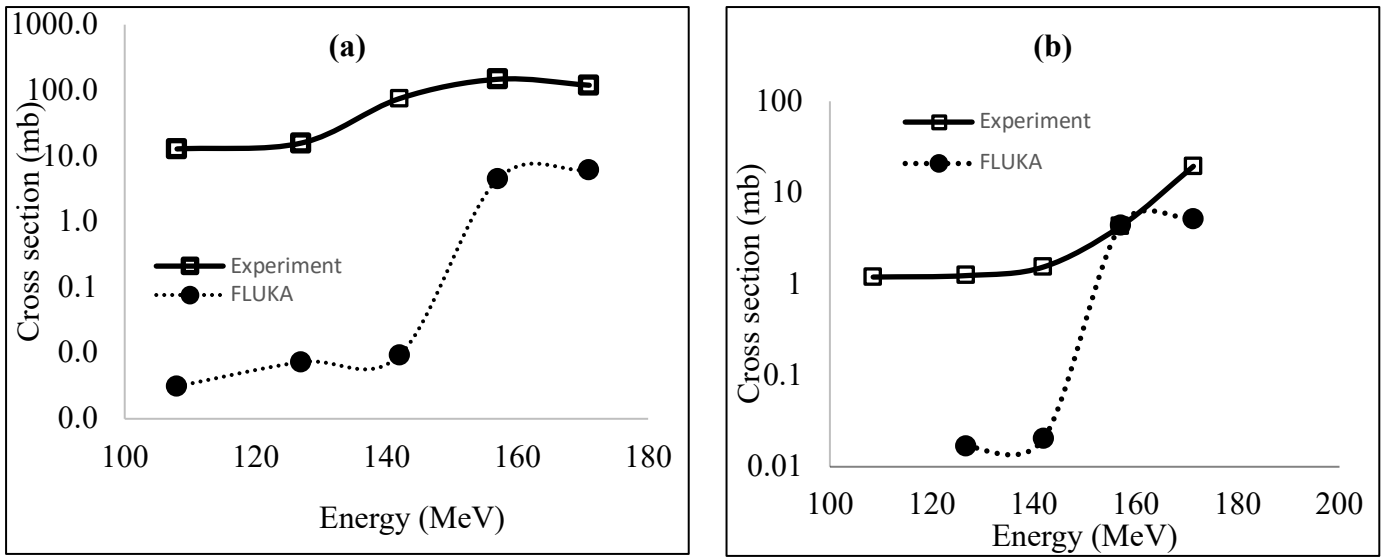

Figure 2: Experimental excitation functions of (a) $^{196}$Au (b) $^{198}$Au and compared with FLUKA

We also found signature of $^{199}$Tl, $^{200}$Tl and $^{201}$Tl which might be produced by multinucleon transfer from the projectile to target. For example, $^{199}$Tl might be produced by 2p transfer from projectile, $^{200}$Tl $^{3}$He transfer from projectile and $^{201}$Tl $^{4}$He transfer from projectile. Both PACE4 and FLUKA nominally announces the presence of these radionuclides in the matrix but cross sections predicted by simulation codes are underpredicted by three to four orders of magnitude compared to the experimental data. The experimental excitations functions of these isotopes have been presented in Figure 3. The production cross-sections of both $^{199}$Tl and $^{200}$Tl are maximum at 157.1 MeV, while that of $^{201}$Tl is maximum at 169 MeV.

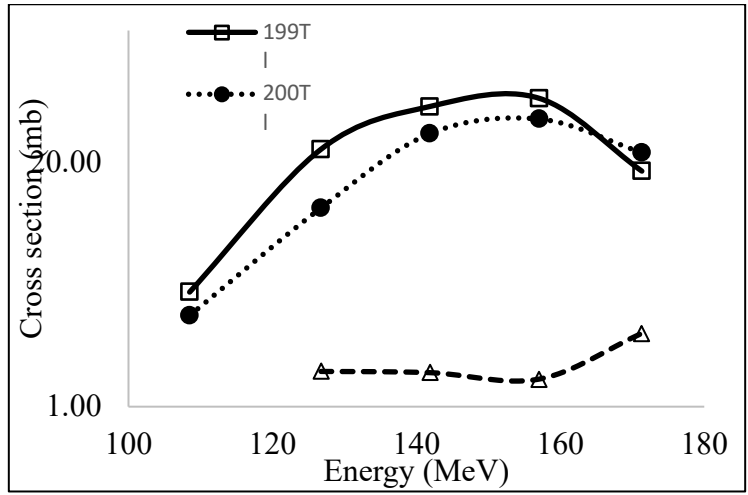

Figure 3: Experimental excitation functions of $^{199}$Tl, $^{200}$Tl, $^{201}$Tl

Table 2 Nuclear characteristics of the radioisotopes identified as direct reaction products.

| Radio-isotope | Half life | Decay mode | Principal gamma lines, keV (Intensity, %) | E $_{\beta^-\text{max}}$, keV (Intensity, %) | Principal Auger electron, keV (Intensity%) |
|---|---|---|---|---|---|
| $^{196}$Au | 6.16 d | ε+β$^+$ (93%) β$^-$ (7%) | 333 (22.9) 355.7 (87.0) 426.1 (6.6) | 261 (6.8) | 7.2 (53.7) |
| $^{198}$Au | 2.69 d | β$^-$ (100%) | 411.8 (95.6) | 961.1 (98.9) | |
| $^{199}$Tl | 7.42 h | ε+β$^+$ (100%) | 208.2 (12.3) 247.2 (9.3) 455.5 (12.4) | | 7.6 (68) |
| $^{200}$Tl | 26.1 h | ε+β$^+$ (100%) | 367.9 (88.5) 579.3 (14.0) 1205.7 (30.4) 1514.9 (4.1) | | 7.6 (61.6) |
| $^{201}$Tl | 3.0 d | ε (100%) | 167.4 (10.0) | | 7.6 (76.8) |

The nuclear characteristics of these radio isotopes have been tabulated in Table 2. Among these $^{198}$Au, due to its moderate half-life and high β$^-$ energy, may serve as potential therapeutic radioisotopes. On the other hand, $^{196}$Au, due to its moderate half-life and high intensity gamma energy may be used as SPECT radio nuclide. Therefore $^{196}$Au-$^{198}$Au would serve as theragnostic pair of radionuclides. Similarly, $^{200}$Tl, a K analog, is also an attractive SPECT radionuclide. All the radionuclides except $^{198}$Au have Auger electrons with good intensity.

# 6 Conclusion

This work will enrich the nuclear data of direct reaction products by the interaction of $^{20}$Ne beam with gold target in the intermediate energy range. Mismatch between the prediction of FLUKA codes and experiments will definitely help to modify the modality of FLUKA code in this energy region. This paper will also help to plan production of direct reaction products, many of which are clinically important, in scale-up mode.

## Acknowledgement

Authors are thankful to target laboratory and cyclotron operating crews of Variable Energy Cyclotron Centre for their help and cooperation. SM is thankful to Department of Science and Technology-INSPIRE Fellowship Program. SL is thankful to Council of Scientific and Industrial Research for providing CSIR Emeritus fellowship.